\documentclass[12pt]{article}
\usepackage[cp1251]{inputenc}
\usepackage{inputenc}
\usepackage{amssymb,amsmath}
\inputencoding{cp1251}
\begin{document}
 \begin{center} {\bf QUANTUM PROTOCOLS AT PRESENCE OF NON-ABELIAN SUPERSELECTION RULES IN THE FRAMEWORK OF ALGEBRAIC MODEL}
\end{center}
\begin{center}A.S.~Sitdikov\footnote{Kazan State Power Engineering University, Kazan, Russia}$^,$\footnote{Kazan (Volga region) Federal University, Kazan, Russia},
A.S.~Nikitin$^1$
\end{center}
\begin{abstract}
In this paper, we study the influence of non-abelian superselection rules on the transfer
of quantum information with the help of qubits on the base of an algebraic model
and formulate quantum protocols. We pay the main attention
to the superselection structure of the algebra of observables $O_G$ defined by
the Cuntz algebra $O_d$ (a field algebra) that contains $O_G$ as
a pointwise fixed subalgebra with respect to the action of the gauge
group $G$. We prove that it is possible to code information only with the help of
states such that projectors on them belong to the algebra of observables
and, owing to their commutativity with elements of the representation of the group $G$,
they allow the recipient to restore the obtained information.
\end{abstract}
\section{Introduction}

In recent time, due to the progress of experimental technology,
the practical implementation of quantum communication, i.e.,
the transfer of information encoded with the help of quantum states of elementary particles,
moves to a new level of development [1--3].

As appeared, the superselection phenomenon plays an important role in studying the quantum information transfer [4, 5].
In nature, there can take place only coherent superpositions of states that correspond to one and the same
eigenvalue of the charge superselection operator, while superselection rules [6] prohibit the occurrence of
superposition states that correspond to its distinct values. Any superpositions of pure states in different
sectors lead to mixed states described by the density matrix.

In the paper [7], we propose an algebraic model for studying
few-nucleon systems with non-abelian superselection rules. The goal
of this paper consists in the application of this model for describing the quantum information transfer
with the help of nucleons at the presence of non-abelian isospin superselection rules.

\section{Preliminary information}
\subsection{Superselection sectors}
The existence of  ``superselection rules'' was first mentioned in the paper~[6].
Ibid, G.C.~Wick, A.S.~Wightman, and E.P.~Wigner give its mathematical statement,
which consists in the existence of orthogonal subspaces $\mathcal{H}_i$ of the full space of states $\mathcal{H}$: $\mathcal{H}=\oplus\mathcal{H}_i$. It is assumed that probabilities of transitions between different orthogonal subspaces
under the action of operators of observables equal zero,
while superpositions of vectors from different such subspaces form mixed states.
Correspondingly, (coherent) superpositions of vectors in given $\mathcal{H}_i$ correspond to pure states;
for this reason, subspaces $\mathcal{H}_i$ are called coherent superselection sectors.

The necessary condition for the representability of $\mathcal{H}$ as the direct sum
of coherent subspaces is the existence of the so-called superselection
operator $\mathfrak{S}$ that commutes with operators of all observables without exception
(including the Hamiltonian of the system). The commutativity with the Hamiltonian
also implies that the superselection operator defines some
absolutely conserved quantity; it is called a superselection charge.
The superselection operator $\mathfrak{S}$ defines a certain superselection rule,
while coherent superselection sectors represent its proper subspaces indexed by its eigenvalues.
The superselection operator is representable as the superposition
\begin{equation}
\mathfrak{S}=\Sigma\mu_i\Pi_i,
\end{equation}
where $\Pi_i$ are projectors on coherent subspaces $\mathcal{H}_i$,
while $\mu_i$ are certain real values~[8]. Note also that $\mathfrak{S}$ belongs to the center
of the algebra of observables $Z(\mathcal{A})$. If the algebra of observables of a physical system
is defined with the help of an abstract $C^*$-algebra $\mathcal{A}$~[9], then, owing to the operator $\mathfrak{S}$,
representations of this algebra $\pi(\mathcal{A})$ in subspaces $\mathcal{H}_i$
are factorial of type $I$ and pairwise disjoint~[10]. In other words,
representations of $\pi(\mathcal{A})$ in subspaces $\mathcal{H}_i$ are irreducible and unitarily equivalent.
Therefore superselection rules prohibit any transfer from some superselection sector to some other one, i.e.,
$<\psi_i|\pi(A)|\psi_j>=0$, where $\psi_i\in\mathcal{H}_i$,
$\psi_j\in\mathcal{H}_j$, $\pi(A)\in\pi(\mathcal{A})$, $A\in\mathcal{A}$.
Therefore, operators of observables on the whole space $\mathcal{H}$ act reducibly.

Though superselection rules began to be taken into account in Physics of Particles
immediately after the publication of the paper~[6] that was mentioned above, the algebraic statement
of the quantum field theory [11--15] has given them a special status.
Therefore these rules have become one of the main tools of any theory of elementary particles.
But nevertheless, the role of superselection rules in establishing the connection between the field algebra
and the algebra of local observables was still not fully understood.
A decisive role here was played by papers [13--15], whose authors have stated the Doplicher--Roberts duality theorem.
From the mathematical point of view, this duality is a generalization of the Tannaka--Krein duality principle
for compact non-abelian groups $G$~[16].
The essence of the generalization consists in the fact that any abstract symmetric
tensor category allows a (functorial) embedding in the category of
Hilbert spaces. Here the symmetry plays a decisive role, because
if a category possesses the symmetry, then the embedding functor exists automatically;
this fact was proved by S.\,Doplicher and J.E.\,Roberts~[14].

From the physical point of view, the duality is connected with the existence
of the gauge groups that play a crucial role in theories of quantum and
quantum field systems. If we restrict ourselves to describing a system
on the base of a field algebra~[15], then the group $G$ of gauge transforms
generates the group of automorphisms of the field algebra $aut(\mathfrak{F})$ and,
consequently, we get the $C^*$-dynamic system $(\mathfrak{F}, G, \alpha).$
Here $\mathfrak{F}$ is a field $C^*$-algebra, $\alpha:$ $g\rightarrow\alpha_g$ $\in aut(\mathfrak{F})$, $g\in G$.
Then we can define the algebra of observables $\mathcal{A}$ as
the invariant part of the field algebra with respect to the action of the gauge group.

If instead of the group $G$ we proceed from the dual object, i.e., the category of its representations,
then we come to the notion of the superselection structure (the totality of superselection sectors)
of the considered physical system and to dynamical superselection rules connected with internal symmetries.
In the case of the abelian group $G\cong U(1)$, its representation group
(the dual group, i.e., the group of characters $\chi(G)$) is isomorphic to the additive group $\mathbb{Z}$
and corresponds to the superselection structure of additive abelian charges (electric, lepton, and
baryonic charges). In the abelian case, the duality is also called the Pontryagin duality.

The dual object of the non-abelian compact group $G$ is the tensor
symmetric category of representations~$\mathbf{rep}(G)$. Moreover, in the paper~[15],
S.\,Doplicher and  J.E. \,Roberts prove the isomorphism of this category with the category of  representations
of the algebra of observables $\mathbf{rep}(\mathcal{A})$
(provided that representations satisfy
the so-called Doplicher--Haar--Roberts selection criterion, which among all possible observables selects
only those that are of physical interest)
that describes the superselection structure of a quantum physical system.
Any representation that satisfies this selection criterion is unitarily
equivalent to the representation of $\mathcal{A}$ in the vacuum Hilbert space $\mathcal{H}_0$
and takes the form $\pi=\pi_0\circ\rho$, where $\rho$ is a localized endomorphism of the alegbra $\mathcal{A}$ [8, 9].
According to the paper~[14], such endomorphisms of the algebra of observables also form the tensor
$C^*$-category ${\bf end(\mathcal{A})}$ which is isomorphic to categories
$\mathbf{rep}(\mathcal{A})$ and $\mathbf{rep}(G)$. Therefore, in studying non-abelian superselection rules,
most important is studying endomorphisms of the algebra of observables,
rather than its automorphisms.

Note also that the base of this duality is the structure of a certain $C^*$-algebra $\mathcal{O}_{\rho}$,
which is called the Doplicher--Roberts algebra (functorially) associated with the object $\rho$
of the tensor symmetric $C^*$-category~[14].

In this paper, we restrict ourselves to considering the case of the strict symmetric tensor
$C^*$-category of finite-dimensional continuous unitary representations of the compact
Lie group ($G=SU(2)$) generated by tensor degrees of the fundamental representation $\pi^{1/2}$
(here the superscript $1/2$ indicates the value of the isotopic spin which we regard as a qubit).
In this case, the base of the duality is the Cuntz $C^*$-algebra $\mathcal{O}_d$~[17] generated
by isometric operators $\psi_i$ ($i=1,2,...,d$) such that
\begin{equation}  \label{kir1}
\psi^*_i\psi_j=\delta_{ij}I
\end{equation}
and
\begin{equation}  \label{kir2}
\sum_i^{d}\psi_i\psi^*_i=I,
\end{equation}
where $I$ is the unit of the Cuntz algebra.

The linear span of these isometries forms the so-called canonical Hilbert space.
The scalar product in such a complex Hilbert space $\mathcal{H}$ of dimension dim\hspace{0.1cm}$d\geq 2$
with an orthonormal basis $\{\psi_i\}_{i=1,2,...,d}$ obeys the formula
\begin{equation}  \label{kir3}
\psi^*\psi'=\langle \psi,\psi'\rangle I,\hspace{0.3cm}\psi,\psi'\in \mathcal{H}.
\end{equation}

In the Cuntz algebra, we can consider Hilbert subspaces of the indicated type,
where products of their intertwining operators are scalar operators [13, 17].
One can identify the tensor product of such subspaces with
the elementwise product of these subspaces in the Cuntz algebra.
Therefore, the category of finite-dimensional Hilbert spaces is
equivalent to the category of finite-dimensional Hilbert subspaces of the Cuntz algebra.

\subsection{The quantum bit and quantum information}

The classical information bit corresponds to one digit in the binary code
that takes on only one of mutually exclusive values (``0'' or ``1'',  `No'' or ``Yes'', etc).
As for the quantum information, its unit is a quantum bit, or a qubit, for short;
it represents a two-level quantum system
and is described by a two-dimensional Hilbert state space.
In other words, a qubit represents the superposition of two orthogonal states
(the superposition of states of a two-level system)
\[
|\psi>=\alpha\binom 10+\beta\dbinom 01.
\]

In quantum mechanics, the measured qubit state takes on this or that possible value with the corresponding probability.
Therefore, a two-level system can carry one bit of classical information.
See, for example, papers [18, 19] for more detail.

The Hilbert space of states of two qubits is four-dimensional,
its basis is the tensor product of bases of separate qubits.

Below in Section~4, we  consider the quantum information transfer
subject to non-abelian superselection rules and see that the latter impose certain constraints.
For example, by decomposing the state space in the orthogonal sum of coherent subspaces
indexed by eigenvalues of the superselection operator,
with the help of two qubits we can transfer only one bit of classical information.

\section{The model}

Since one usually studies the behavior of a certain physical quantum system
on the base of measurement procedures, it makes sense to describe this system
on the base of the algebra of physical observables $\mathcal{A}.$
Then the analysis of measurement results is reduced to the summation and multiplication of operators
that correspond to the observed values.
This naturally leads to internal associative composition rules with semigroup properties.
Moreover, according to results obtained in papers~[13, 15], the algebra of such observables
can be consistently embedded in another algebra, namely,
the extended one obtained by means of the cross product $\mathcal{A}\times\mathcal{T}$, where $\mathcal{T}$
is a complete subcategory $\mathcal{T}\subset{\bf end(\mathcal{A})}$
\footnote{A certain part of the material in this section is connected with notions and denotations introduced in Subsection~2.1,
therefore we use them without references.}.
The mentioned cross product corresponds to the field algebra $\mathfrak{F}$.
Its construction actually justifies the possibility of obtaining the dynamic system $(\mathfrak{F}, G, \alpha)$
from a given abstract category of endomorphisms in the sense that this abstract category
is the dual object of the compact group $G$ instead of the concrete category $\mathbf{rep}(G)$.
Without going into detail (see, if necessary, papers~[13, 15]), note that the group of authomorphisms
in the $C^*$-algebra $\mathcal{A}\times\mathcal{T}$ constructed in the mentioned way is the group $G$,
while $\mathcal{A}\times\mathcal{T}$ contains the $C^*$-algebra $\mathcal{A}$ as its pointwise
fixed sublgebra with respect to the action of this group.

If we restrict ourselves to studying the subcategory $\mathcal{T}_{\rho}\subset\mathcal{T}$ generated by tensor degrees
of one endomorphism $\rho$, provided that morphisms of this subcategory generate the $C^*$-algebra $\mathcal{A}$,
then (see~[13]) $\mathcal{A}\times\mathcal{T}_{\rho}$ represents nothing else but the Cuntz algebra $O_d$,
where $d$ is the dimension of the endomorphism $\rho$ generated
by the fundamental multiplet (isometries) $\psi_i$ ($i=1,2,...,d$) satisfying correlations~(2) and~(3).
Therefore, in this case, instead of intertwining operators (morphisms in the category of endomorphisms),
we can consider maps between Hilbert spaces formed by these isometries
and describe sectors with the help of decompositions of degrees of the canonical Hilbert space (see below),
where there act irreducible representations of the group $SU(d)$
\footnote{In this case, the duality is described in terms of Lie groups.}.

In this paper, we study the quantum information transfer with the help of particles with the isospin of $T=1/2,$
therefore we restrict ourselves to considering the non-abelian gauge group $G=SU(2)$ of isotopic rotations.
As an orthonormal basis of the Hilbert space $\mathcal{H}_{1/2},$
where the fundamental representation $\pi^{1/2}$ of the group $G=SU(2)$ takes place,
we choose the multiplet $\psi_i$ $(i=1,2)$ that satisfies correlations~(2)--(3).
Below, when this leads to no misunderstanding, for simplicity, we omit the index $1/2$
and just write $\mathcal{H}$ and $\pi$. Such a multiplet generates the $*$-algebra $^0\mathcal{O}_2$, whose completion
with respect to the $C^*$-norm forms the Cuntz algebra $\mathcal{O}_2$. Its subalgebra $O_{G=SU(2)},$
which remains pointwise fixed with respect to the action of the group $SU(2)$,
is associated with the algebra of observables.

As is proved in papers~[13, 17], generators of this algebra with arbitrary $d$ are operators
\begin{equation}\vartheta
(p_n)=\sum_{i_1i_2...i_n}\psi_{i_1}...\psi_{i_n}\psi^*_{i_{p(n)}}...\psi^*_{i_{p(1)}},
\end{equation}
where $p_n\in\mathbf{P}_{n}$ and
\begin{equation}S=\frac{1}{\sqrt{d!} }\sum_{p\in\mathbf{P}_{d}}sign(p)\psi
_{p(1)} ... \psi _{p(d)}.
\end{equation}
Here $\mathbf{P}_n$ is a symmetric group, $\mathbf{P}_d\subset\mathbf{P}_n$
(for brevity, hereinafter we omit symbols of the tensor product in $\psi_i\otimes\psi_j$ and mainly use these denotations).

Superselection sectors are labeled here by isospin quantum numbers $T=0; \frac{1}{2}; 1; \frac{3}{2};...$
and each such subspace (sector) $\mathcal{H}_T$ is indexed with the help of a fixed value of the isospin $T$
which is characterized by various values of its projections $T_z$, whose number (it equals $2T+1$)
defines the dimension of the sector.
The linear span formed by isometries $\psi_{i_1}\psi_{i_2}...\psi_{i_r}$ forms the $r$th tensor degree of the $d$-dimensional space
$\mathcal{H}_{T}:$ $\underbrace{\mathcal{H}_{T}\otimes...\otimes\mathcal{H}_{T}}_r;$ we denote it as $\mathcal{H}^r$.
Tensor degrees $\pi^r$ of the representation $\pi$ act as unitary operators in $\mathcal{H}^r$.
This representation is reducible; it decomposes in the direct sum of irreducible representations (superselection sectors)
that act on proper coherent subspaces of the Hilbert space $\mathcal{H}^r$.

Linear maps $\mathcal{H}^s\rightarrow \mathcal{H}^r$ $\equiv(\mathcal{H}^s, \mathcal{H}^r)$ between tensor degrees
form a linear span of combinations in the form
\begin{eqnarray}
\psi_{i_1}...\psi_{i_r}\psi^*_{j_s}...\psi^*_{j_1}\in(\mathcal{H}^s, \mathcal{H}^r).
\end{eqnarray}
Spaces of $G$-invariant intertwining operators $(\mathcal{H}^r,\mathcal{H}^s)_{SU(d=2)}$
generate an algebra of observables, i.e., a pointwise fixed subalgebra of the Cuntz algebra
whose generators are operators in form~(5),~(6).

In the case when $d=2$, formula~(6) implies that
\begin{eqnarray}
S=\frac{1}{\sqrt{2}}(\psi_1\psi_2-\psi_2\psi_1),
\end{eqnarray}
and $SS^*$ represents a projector on the completely antisymmetric subspace
in the tensor product $\mathcal{H}^{r=2}=\mathcal{H}\otimes\mathcal{H}$.
Here coherent proper subspaces are the one-dimensional space $\mathcal{H}_0$ and the three-dimensional one
$\mathcal{H}_1,$
\begin{eqnarray}\mathcal{H}^2=\mathcal{H}_0\oplus\mathcal{H}_1.
\end{eqnarray}
Bases of these spaces are obtained by representing the tensor $\psi_i\psi_j$ $(i,j=1,2)$
as the sum of the antisymmetric and symmetric tensors, correspondingly:
\footnote{The summation of isospins of two particles with $T=1/2$ leads to the full isospin either with $T=0$
or with $T=1$ and $T_z=1,0,-1$, therefore, in this case, for convenience, we use double subscripts for isometric operators.}
\begin{equation}  \label{kir7}
\psi_{00}=\frac{1}{\sqrt{2}}(\psi_1\psi_2-\psi_2\psi_1)
\end{equation}
and
\begin{equation}  \label{kir8}
\psi_{11}=\psi_1\psi_1,\hspace{0.2cm}
\psi_{10}=\frac{1}{\sqrt{2}}(\psi_1\psi_2+\psi_2\psi_1),\hspace{0.2cm}
\psi_{1-1}=\psi_2\psi_2.
\end{equation}
Respectively, the representation matrix decomposes in the dirtect sum
of the trivial ($T=0$) and vector ($T=1$) representations $\pi^2=\pi\otimes\pi=\pi_0\oplus\pi_1$.
Expression~(10) (as it should be) coincides with~(8) and the projector $\Pi_{00}$
on the completely antisymmetric one-dimensional subspace in $\mathcal{H}^2$ is
\begin{eqnarray}
\Pi_{00}=\psi_{00}\psi^*_{00}=SS^*.
\end{eqnarray}
Analogously, denote projectors on basis states~(11) as
\begin{equation}
\Pi_{11}=\psi_{11}\psi^*_{11}=\psi_1\psi_1\psi^*_1\psi^*_1;
\end{equation}
\begin{equation}
\Pi_{10}=\psi_{10}\psi^*_{10}=\frac{1}{2}(\psi_1\psi_2+\psi_2\psi_1)(\psi^*_2\psi^*_1+\psi^*_1\psi^*_2);
\end{equation}
\begin{equation}
\Pi_{1-1}=\psi_{1-1}\psi^*_{1-1}=\psi_2\psi_2\psi^*_2\psi^*_2.
\end{equation}

Let us now make some mathematical remarks about algebras $O_G$
that are obtained from the category of tensor degrees of a finite-dimensional
Hilbert space and have the same mathematical structure
as the Doplicher--Roberts $C^*$-algebra $\mathcal{O}_{\rho}.$

Tensor degrees
\footnote{We consider the case of $d=2.$}
$\mathcal{H}^{r}$ ($r=0,1,2,...$)
of the canonical Hilbert space $\mathcal{H},$ in general, form a subcategory in the category of Hilbert spaces
${\bf hilb}$, where morphisms between degrees $\mathcal{H}^{s}$ and $\mathcal{H}^{r}$ are maps in form~ (7).
Objects in this subcategory form $G$-modules, where there act continuous unitary representations of the group $SU(2)$
that form the monoidal category ${\bf rep(G)}$ of tensor degrees of the irreducible representation $\pi$
of the compact group $SU(2).$
Its objects are
\begin{equation*}
Obj {\bf rep(G)} = \{\iota, \pi, \pi^{2},
\pi^{3},...,\pi^{
r},...,\pi^{s},...\},
\end{equation*}
where $\pi\otimes\pi\equiv \pi^{2},...$; $r,s\in\mathbf{N}_0$.
Morphisms in this category are intertwining operators of such representations.
Correspondingly, in this case, morphisms are $G$-module homomorphisms ($\mathcal{H}_{T}^r,\mathcal{H}_{T}^s)_G$.
This structure allows us to define the subalgebra $^{0}\mathcal{O}_{G}\subset$ $^{0}\mathcal{O}_2$ of the Cuntz algebra
as follows. Taking into account~(3) and~(7),
we can easily make sure that
$(\mathcal{H}_{T}^{r},\mathcal{H}_{T}^{r+k})_G\longrightarrow^{\otimes
1}(\mathcal{H}_{T}^{r+1},\mathcal{H}_{T}^{r+1+k})_G$
is an injective map of morphisms.
Therefore, the inductive limit of the sequence
\begin{equation*}
(\mathcal{H}_{T}^{r},\mathcal{H}_{T}^{r+k})_G\longrightarrow^{\otimes
1}(\mathcal{H}_{T}^{r+1},\mathcal{H}_{T}^{r+1+k})_G\longrightarrow^{\otimes
1}...\longrightarrow^{\otimes
1}(\mathcal{H}_{T}^{r+n},\mathcal{H}_{T}^{r+n+k})_G\longrightarrow^{\otimes
1}...,
\end{equation*}
with fixed $k$ defines some Banach space $^{0}\mathcal{O}^k_G$
(the composition of morphisms is defined here in a natural way).
Then the summation over all $k$ gives the $\mathbb{Z}$-graded $C^*$-algebra
$^{0}\mathcal{O}_G=\bigoplus_{k\in\mathbf{Z}}^{0}\mathcal{O}^k_G$
with a nontrivial endomorphism in the form
$\sigma_{\mathcal{H}}=\sum_{i=1}^{d=2}\psi_iX\psi^*_i$,\hspace{0.2cm}$X\in O_{SU(2)},$ acting in it.
The closure of this algebra $^0\bar{\mathcal{O}_G}$ leads to the subalgebra $\mathcal{O}_G$ of the Cuntz algebra.

Tensor degrees of the endomorphism $\sigma_{\mathcal{H}}$ form a category, namely,
the category of endomorphisms of the algebra of observables;
in the Doplicher--Roberts approach, just they define the superselection structure of the theory.
However, one can easily make sure that morphisms $t\in(\sigma^s_{\mathcal{H}},\sigma^r_{\mathcal{H}})$ in this category are
the same maps in form~(7). One can also easily prove that tensor degrees are reducible and decompose in the direct sum of irreducible ones. For example, in the case of a tensor square, we get the decomposition
\begin{equation}\sigma^2_{\mathcal{H}}=\sigma_{\mathcal{H}\otimes\mathcal{H}}(X)=\sigma^0(X)\oplus\sigma^1(X),
\end{equation}
where $$\sigma^0(X)=\psi_{00}X\psi^*_{00},\hspace{0,2cm} \sigma^1(X)=\psi_{11}X\psi^*_{11}+\psi_{10}X\psi^*_{10}+\psi_{1-1}X\psi^*_{1-1},\hspace{0,2cm}X\in O_{SU(2)}.$$

The symmetry in this category is defined by unitary operators in the form
$$\vartheta(r,s)=\mathcal{H}^r\otimes\mathcal{H}^s \rightarrow \mathcal{H}^s\otimes\mathcal{H}^r;$$
each of these operators corresponds to a concrete permutation
$$\left(\begin{array}{cccccccc}
1&2&...&r&r+1&r+2&...& r+s \\
s+1&s+2&...&s+r&1&2&...& s \\
\end{array}\right).$$
With $d=2$ and $r=s=1$, using formula~ (5), we get the correlation
$$\vartheta(1,1)=\vartheta=\sum_{i_1i_2}^2\psi_{i_1}\psi_{i_2}\psi^*_{i_{p(2)}}\psi^*_{i_{p(1)}}=\psi_1\psi_1\psi^*_1\psi^*_1
+\psi_1\psi_2\psi^*_1\psi^*_2+\psi_2\psi_1\psi^*_2\psi^*_1+\psi_2\psi_2\psi^*_2\psi^*_2.$$

The operator $S$ in this category defines the determinant of a special object~[14];
in this model, this object is $\sigma_{\mathcal{\mathcal{H}}}.$

Therefore, in our case, from the formal mathematical point of view,
descriptions of sectors both with the help of the category of tensor degrees
of the endomorphism $\sigma_{\mathcal{H}}$ and with the help of the category of representations of the group
$SU(2)$ are equivalent .

\section{The quantum transfer of isospin states}

Basing on the proposed model, let us consider the information transfer with the help of qubitis.
A particle with the isospin $T=1/2$ and its projections $T_z=\pm 1/2$ represents
a two-level quantum system (a qubit).
In our model, states of such a particle are described by the two-dimensinal Hilbert space $\mathcal{H}$,
whose orthonormal basis is formed by isometries $\psi_1$ and $\psi_2$.
Assume that Alice wants to send some message to Bob with the help of a pair of particles (two qubits).
The Hilbert space of states of such a system is formed by coherent subspaces~(9),
where $\mathcal{H}_0$ corresponds to the state with the full isospin $T=0$,
and  $\mathcal{H}_1$ does to the state with $T=1$.
At the same time, the projector on $\mathcal{H}_0$ obeys expression~(12),
while the projector on the whole subspace $\mathcal{H}_1$ is defined by the sum
of projectors~(13)--(15); we denote it as $\Pi_1=\Pi_{11}+\Pi_{10}+\Pi_{1-1}.$
Hence it also follows that $\Pi_{00}=\sigma^0(I)=1_{\mathcal{H}_0}$ and $\Pi_{1}=\sigma^1(I)=1_{\mathcal{H}_1}.$
Here $I$ is the unit of the Cuntz algebra,
$1_{\mathcal{H}_0}$ and $1_{\mathcal{H}_1}$ are identical operators in these subspaces.
However, projectors $\Pi_{1i}$ ($i=1,0,-1$) on subspaces with distinct $T_z$
depend on the orientation of the system of coordinates in the isospin space;
in this sense they do not relate to the algebra of observables $O_G$,
which by definition represents a pointwise fixed subalgebra with respect to the action of the group $G.$
Really, the fundamental representation of the group $SU(2)$ that acts on the Hilbert space $\mathcal{H}$
is defined by the unitary unimodular matrix
\begin{eqnarray}
\pi(g)=\left(\begin{array}{cc} {\alpha } & {\beta } \\
{-\bar{\beta}} & {\bar{\alpha}} \end{array}\right),
\end{eqnarray}
whose matrix elements are such that
$|\alpha|^2+|\beta|^2=1$ and $\alpha\bar{\alpha}+\beta\bar{\beta}=1.$
It is convenient to parameterize these matrix elements as follows:
\begin{eqnarray}
\alpha=\cos\frac{\theta}{2}\exp\left(i\frac{\varphi_1+\varphi_2}{2}\right);
\nonumber\\
\beta=i\sin\frac{\theta}{2}\exp\left(-i\frac{\varphi_2-\varphi_1}{2}\right);
\nonumber\\
\nonumber\\
\bar{\alpha}=\cos\frac{\theta}{2}\exp\left(-i\frac{\varphi_1+\varphi_2}{2}\right);
\nonumber\\
\bar{\beta}=-i\sin\frac{\theta}{2}\exp\left(i\frac{\varphi_2-\varphi_1}{2}\right).
\end{eqnarray}
Here $\varphi_1,\varphi_2, \theta$ are Euler angles such that $0\leq\varphi_1\leq 2\pi$, $0\leq\varphi_2\leq 2\pi$, $0\leq\theta\leq \pi.$
With respect to the system of coordinates characterized by Euler angles $\varphi_1=\varphi_2=\theta=0,$
the state that corresponds to the projection $T_z=1$ is the basis state $\psi_{11}=\psi_1\otimes\psi_1.$
With respect to the rotated coordinate system, for this state we get the formula
$$\pi(g)(\psi_1\otimes\psi_1)=\alpha^2\psi_{11}-\sqrt{2}\alpha\bar{\beta}\psi_{10}+\bar{\beta}^2\psi_{22},$$
i.e., it turns into the coherent superposition of basis vectors of the space $\mathcal{H}_{1}$
(for clarity, we preserve the symbol of the tensor product).
The projector $\Pi_{11}$ on the state $\psi_1\otimes\psi_1$ then takes the form
\begin{eqnarray}
\pi(g)\Pi_{11}\pi^+(g)
=(\alpha\bar{\alpha})^2\Pi_{11}+2(\alpha\bar{\alpha})(\beta\bar{\beta})\Pi_{10}+
(\beta\bar{\beta})^2\Pi_{1-1}+
\alpha^2\beta^2\psi_{11}\psi^*_{1-1}+\nonumber\\+\bar{\alpha}^2\bar{\beta}^2\psi_{1-1}\psi^*_{11}-
\sqrt{2}(\alpha^2\bar{\alpha}\beta\psi_{11}\psi^*_{10}+\bar{\alpha}^2\alpha\bar{\beta}\psi_{10}\psi^*_{11}+\alpha
\bar{\beta}\beta^2\psi_{10}\psi^*_{1-1}+\nonumber\\+\bar{\alpha}\beta\bar{\beta}^2\psi_{1-1}\psi^*_{10}).
\end{eqnarray}
Here $\pi^+(g)$ is the conjugate unitary operator.

Assume now that Alice sends information in the symmetric state $\psi_{11}$
prepared in her coordinate system with the help of the projector $\Pi_{11}.$ Since Bob,
at the same time, is not aware of the Alice's system of coordinates, he has
to perform the averaging with respect to the whole group, namely,
\begin{equation}
\tilde{\Pi}_{11}=\int_G\pi(g)\Pi_{11}\pi^+(g)d\mu(g).
\end{equation}
Here $d\mu(g)=(1/8\pi^2)\sin\theta d\theta d\varphi_1d\varphi_2$ is an invariant Haar measure.
For getting an explicit form of $\tilde{\Pi}_{11}$ by formula~(20), we have to integrate expression~(19);
actually, this implies the integration of coefficients at projectors. For example, in accordance with formula~(18),
for the mean value $\widehat{(\alpha\bar{\alpha})}^2$ we get the correlation
$$\widehat{(\alpha\bar{\alpha})}^2=\frac{1}{8\pi^2}\int^{\pi}_{0}\cos^4\frac{\theta}{2}\sin\theta d\theta\int^{2\pi}_{0}d\varphi_1
\int^{2\pi}_{0}d\varphi_2=\frac{1}{3}.$$
Integrating analogously the rest coefficients in formula~(19) and taking into account the fact
that the averaging of coefficients at projectors that map basis states in $\mathcal{H}_{1}$ to each other gives zero,
we get the formula
$$\tilde{\Pi}_{1}=\frac{1}{3}\left(\Pi_{11}+\Pi_{10}+\Pi_{1-1}\right)=\frac{1}{3}\Pi_1.$$
Now we can easily prove that $[\tilde{\Pi}_{11}, \pi(g)]=0,$ which means that $\tilde{\Pi}_{11}$
belongs to the algebra of observables $O_{G}.$
Therefore, the averaging over the group leads to the decoherence~[20]:
the pure state $\rho=\Pi_{11}$ sent by Alice is treated by Bob as a mixed one $\tilde{\rho}=\tilde{\Pi}_{11}.$
We can say that by using two qubits Alice can send to Bob one classical information bit, following the protocol:
{\em with the help of the antisymmetric state}~(10) {\em send the information that corresponds to $0$ and
with the help of any state}~(11) {\em from the symmetric subspace also send the information that corresponds to $1$} (see item~2.2).
Here Bob performs the projective measurement on antisymmetric and symmetric subspaces with the help of
a superselection operator in the form (see formula~(1))
$$\mathfrak{S}=p_1\left(\frac{1}{3}\Pi_1\right)+p_0\Pi_{00},$$
where $p_1$ and $p_0$ are probabilities defined by frequencies of sent signals encoded by symmetric and antisymmetric states, correspondingly.

In the case of two particles, the projector $\Pi_{00}$ on the antisymmetric subspace is a scalar value,
and with the help of formula~(20) one can easily make sure that $\tilde{\Pi}_{00}=\Pi_{00}$.
Therefore, Alice can code messages only with the help of states such that projectors on them belong to the algebra of observables $O_{SU(2)}$. Owing to the commutativity of such projectors with elements $\pi(g)$, Bob unambiguously recognizes the received information.

Let us now consider the transfer of the quantum information with the help of three qubits.
In this case, we get an 8-dimensional reducible Hilbert space.
It decomposes in the direct sum of coherent subspaces:
$\mathcal{H}^3=Lin\{\psi_i\psi_j\psi_k\}_{i,j,k=1}^{d=2}=
2\mathcal{H}_{\frac{1}{2}}\oplus\mathcal{H}_{\frac{3}{2}},$
whose bases consist of the following tensors:

\begin{eqnarray}
e_1=\frac{1}{\sqrt{2}}\left(\psi_1\psi_2\psi_1-\psi_2\psi_1\psi_1\right)\nonumber\\
e_2=\frac{1}{\sqrt{2}}\left(\psi_1\psi_2\psi_2-\psi_2\psi_1\psi_2\right),
\end{eqnarray}

\begin{eqnarray}
e_3=\frac{1}{\sqrt{6}}\left(2\psi_1\psi_1\psi_2-\psi_1\psi_2\psi_1
-\psi_2\psi_1\psi_1\right),\nonumber\\
e_4=\frac{1}{\sqrt{6}}\left(\psi_1\psi_2\psi_2+\psi_2\psi_1\psi_2-
2\psi_2\psi_2\psi_1\right);
\end{eqnarray}

\begin{eqnarray}
e_5=\psi_1\psi_1\psi_1,\nonumber\\
e_6=\frac{1}{\sqrt{3}}\left(\psi_1\psi_2\psi_1+\psi_2\psi_1\psi_1
+\psi_1\psi_1\psi_2\right),\nonumber\\
e_7=\frac{1}{\sqrt{3}}\left(\psi_1\psi_2\psi_2+\psi_2\psi_1\psi_2+
\psi_2\psi_2\psi_1\right),\nonumber\\
e_8=\psi_2\psi_2\psi_2.
\end{eqnarray}
Here expressions~ (21) and~(22), taking into account the multiplicity, correspond to the value $T=1/2$,
while expression~(23) does to $T=3/2.$
At the same time, the matrix of the representation $\pi^{1/2}\otimes\pi^{1/2}\otimes\pi^{1/2}$
also decomposes in the direct sum of irreducible representations $2\pi^{1/2}\oplus\pi^{3/2}$.
We can express projectors on these subspaces in terms of unitary operators~(5) and isometries~(6) with $d=2$,
i.e., they belong to the algebra of observables $O_{SU(2)}.$
For example, one can easily make sure that $\Pi_{\frac{1}{2}}=e_1e^*_1+e_2e^*_2=SS^*.$
Therefore, with the help of three qubits Alice can send a message to Bob, adhering to the following protocol:
{\em with the help of any state}~(21) or~(22) {\em send the data that correspond to the sector with $T=1/2$ and
with the help of any state}~(23) {\em from the symmetric subspace also send the data that correspond to the sector with $T=3/2$}.
Hence it follows that the number of classical messages (in this case, it equals three)
that can be coded with the help of three qubits
depends on the number of coherent superselection sectors (taking into account their multiplicities).
One can also easily get an explicit dormula for the superselection operator.

Analogously, for four particles, we get the 16-dimensional state space
$\mathcal{H}^4=Lin\{\psi_i\psi_j\psi_k\psi_l\}_{i,j,k,l=1}^{d=2}.$
However, one can find a basis in it,
in which the representation matrix decomposes in the direct sum
$2\pi^{0}\oplus3\pi^{1}\oplus\pi^{2}$ of irreducible representations, where
the multiplicity of the isosinglet representation equals two,
that of the isotriplet representation equials three,
and the multiplicity of the irreducible representation that corresponds to the value $T=2$ equals one.
Let us give here only the expressions for basis vectors that correspond to the isosinglet representation, i.e.,
\begin{equation}
e_1=\frac{1}{2}\left(\psi_1\psi_2\psi_1\psi_2-
\psi_1\psi_2\psi_2\psi_1-\psi_2\psi_1\psi_1\psi_2+
\psi_2\psi_1\psi_2\psi_1\right)
\end{equation}

$$e_2=\frac{\sqrt{3}}{3}(\psi_1\psi_1\psi_2\psi_2-\frac{1}{2}\psi_1\psi_2\psi_1\psi_2-\frac{1}{2}\psi_2\psi_1\psi_1\psi_2-
\frac{1}{2}\psi_1\psi_2\psi_2\psi_1-$$
$$-\frac{1}{2}\psi_2\psi_1\psi_2\psi_1+\psi_2\psi_2\psi_1\psi_1)\eqno{(25)}$$

One can easily make sure that projectors on these states belong to the algebra of observables $O_{SU(2)}.$
Thus, for example, $e_1e_1^*=SSS^*S^*$ and $e_2e_2^*=1/2(I+\vartheta(1,1)).$
In this case, the number of classical messages that can be coded with the help of four qubits
also equals the number of coherent superselection sectors, taking into account their multiplicitices, i.e., six (2+3+1).

\section{Conclusion}

In this paper, we study the formal possibility of the transfer of quantum information with the help of qubits
at the presence of non-abelian MS spin superselection rules on the base of a simple algebraic model
proposed by us in the paper~[7]. Inspite of the similarity with the theory of representation groups
in finite-dimensional Hilbert spaces, in our model, finite-dimensional Hilbert subspaces belong to the Cuntz algebra and form
a tensor symmetric category, whose $G$-module homomorphisms define an algebra of observables.

In this connection, we focus on the superselection structure of the theory
that is directly determined by the algebra of observables $O_G$ with the trivial center $\mathbb{C}I$ and a subcategory of the category of endomorphisms. This approach allows us to immediately operate with observed values and states defined on the algebra of observables.
Moreover, owing to the Doplicher--Roberts duality theorem, we can extend this approach to the case of an abstract $C^*$-algebra of observables; this allows us to calculate the particle statistics, basing only on the statistics of the superselection sector itself, regardless of the field algebra~[15].
The local structure is also defined here in a natural way,
because observables in various superselection sectors mutually commute.
In this regard, the superselection charge refers to the classical (macroscopic) observable, namely,
since it belongs to the center of the algebra of observables, it commutes with observables in any sector.

Note that the main results obtained here correlate with analogous results established in the paper~[5].
From the mathematical point of view, we can assume that by increasing the number of qubits, one can increase
the effectiveness of information transfer, because the increase of their number leads to the increase of the number of coherent suprtselection sectors, owing to the increase of their multiplicity.

However, from the point of view of the observability, we can say that correlations between particles
do no necessarily lead to bound states in all coherent superselection sectors.
This can lead to certain experimental constraints.
For example, the isospin of the thoroughly studied two-nucleon system
can equal either $T=0$ or $T=1$ (the ordinary spin has the same connection scheme). Correspondingly,
these states belong to two distinct coherent superselection sectors.
However, it is known that two nucleons of total isospin $0$,
which belong to an antisymmetric subspace, are most correlated and correspond to the deuteron observed in the experiment.
At the same time, two nucleons in the superselection sector with $T=1$, which correrspond to symmetric states,
can be realized only as unstable formations;
they correspond to two protons, two neutrons, or a symmetric neutron-proton pair.
In other words, the tensor part of the nuclear interaction,
which depends on the mutual orientation of spins of nucleons,
has a great intensity only with the minimum isospin.

Analogously, taking into account the isospin superselection, a three-nucleon system
can form the bound doublet state with the basis~(21) and~(22)
which corresponds to the minimum isospin of the system $T=1/2$.
These states ($e_1,  e_2$ and $e_3, e_4$) of three nucleons with the isospin $T=1/2$ (tritium and helium-3) are formed with
the help of mixed representations of the symmetric group and are stable.
As for states of three neutrons $(3n)$ and three protons $3p=^3Li$,
they correspond to the symmetric representation with $T=3/2$ (basis~(23)) and are unstable
\footnote{
From the physical point of view, such an isodoublet corresponds to two mirror nuclei
with one proton and two neutrons (the tritium nucleus, i.e., deuteron + neutron)
and with one neutron and two protons  (helium-3, i.e. deuteron + proton).}.

From the physical point of view, four-nucleon systems are structures in the form $4n,$ $4p,$ $1n3p,$ etc;
among them there also exists the structure that corresponds to the nucleus of $^4He,$
i.e., an $\alpha$-particle.
According to experimental results, the $\alpha$-particle has zero spin and isospin.
According to our model, there exist two states that correspond to isotopic singlets~(24),~ (25),
which remain pointwise fixed with respect to the group action.
However, the state defined by expression~(24) represents the product of two operators $S$,
which allows us to consider that this state corresponds to the observed $\alpha$-particle, i.e.,
a stable structure of a four-nucleon system.

\begin{center}
References
\end{center}
\begin{description}
\item[1] Naomi H. Nickerson, Joseph F. Fitzsimons, and Simon C. Benjamin,
``Freely Scalable Quantum Technologies Using Cells of 5-to-50 Qubits with Very Lossy and Noisy Photonic Links'',
Phys. Rev. X 4:4, (2014), 041041.
 \item[2] W.~Kobus, W.~Laskowski, T.~Paterek, M.~Wie{\'s}niak, and H.~Weinfurter,
 ``Higher dimensional entanglement without correlations'',
Eur. Phys. J. D 73(2):29 (2019), 29--39.
 \item[3] Gisin, N. et al. ``Quantum communication technology'',
Electronics Letters, 46(14) (2010), 965.
\item[4] A. Kitaev, D. Mayers, J. Preskill, ``Superselection rules and quantum protocols'',
Phys. Rev. A69, (2004), 052326-1.
\item[5] S.D. Bartlett, T. Rudolf, R.W. Spekkens, ``Reference frames, superselection rules,
and quantum information'', Rev. of Mod. Phys., 79 (2007), 555.
\item[6] G.C. Wick, A.S. Wightman, E.P. Wigner,
``The Intrinsic Parity of Elementary Particles'', Phys. Rev., 88 (1952), 101.
\item[7] M.I. Kirillov, A.S. Nikitin,  and A.S. Sitdikov, ``An Algebraic Model of Nucleon Systems with Rules of Isospin Superselection'',
Izv. Ross. Akad. Nauk, Ser. Fiz., 82:10 (2018), pp. 1403--1407.
 \item[8] Horuzhy, S. S. Introduction to Algebraic Quantum Field Theory. Nauka, Moscow, 1986. -- 304 p. [in Russian].
\item[9] R. Haag. Local Quantum Physics, second ed., Texts and Monographs in Physics, Springer Verlag, Berlin, 1996.
 \item[10] Bogolyubov, N.N., Logunov, A.A., Oksak, A.I., and Todorov, I.T.
 General Principles of Quantum Field Theory. Nauka, Moscow, 1987. -- 615 p. [in Russian].
\item[11] S. Doplicher, R. Haag, J.E. Roberts, ``Local observables and particle statistics I'', Comm. Math. Phys., 23 (1971), 199.
\item[12] S. Doplicher, R. Haag, J.E. Roberts, ``Local observables and particle statistics II'', Comm. Math. Phys., 35 (1974), 49.
 \item[13]  Doplicher, S., Roberts, J.E., ``Endomorphisms of C*-algebras, cross products and duality for compact groups'', Annals of Mathematics, 130 (1989), 75.
 \item[14] Doplicher, S., Roberts, J.E., ``A new duality theory for compact groups'', Invent. Math., 98 (1989), 157.
\item[15] Doplicher, S., Roberts, J.E., ``Why there is field
algebra with a compact gauge group describing the superselection structure in particle physics'', Comm. Math. Phys., 131 (1990), 51.
\item[16] T. Tannaka, ``Uber den Dualitatssatz der nichtkommutativen topologichen Gruppen'', Tohoku Math. J. 45 (1939),  1-12.
1; M. G. Krein, ``A principle of duality for a bicompact group and a square block-algebra'', Dokl. Akad. Nauk SSSR, 69 (1949), 725.
\item[17] Doplicher, S., Roberts, J.E., "Duals of compact Lie groups realized in the Cuntz algebras and their actions on C*-algebras", Jour. of Funct. Analysis, 74
(1987), 96.
\item[18] Bouwmeester, Dirk, Ekert, Artur K., Zeilinger, Anton (Eds.). The Physics of Quantum Information. Postmarket, Moscow, 2002. -- 376 p. [Russ. transl.]
\item[19] Nielsen, M.A. and Chang, I. Quantum Computation and Quantum Information.
Mir, Moscow, 2006. - 824 p. [Russ. transl.].
\item[20] M.B. Menskii. Quantum Measurement and Decoherence.
Fizmatlit, Moscow, 2001. --  227 p.
\end{description}
\end{document}